\documentclass{kluwer}
\usepackage{graphicx}

\begin{document}
\begin{article}
\begin{opening}
\title{GENERALIZED REGULARIZATION TECHNIQUES WITH
CONSTRAINTS FOR THE ANALYSIS OF SOLAR BREMSSTRAHLUNG X-RAY
SPECTRA}


\author{Eduard P. \surname{Kontar}\thanks{Off-print requests: {\tt eduard@astro.gla.ac.uk}}}
\institute{Department of Physics \& Astronomy, The University of
Glasgow, G12 8QQ, UK}
\author{Michele \surname{Piana}}
\institute{Dipartimento di Matematica, Universit\`a di Genova, via
Dodecaneso 35, I-16146 Genova, Italy}
\author{Anna Maria \surname{Massone}}
\institute{INFM, UdR di Genova, via Dodecaneso 33, I-16146 Genova,
Italy}
\author{A. Gordon \surname{Emslie}}
\institute{Department of Physics, The University of Alabama in Huntsville, Huntsville, AL 35899,
USA}
\author{John C. \surname{Brown}}
\institute{Department of Physics \& Astronomy, The University of Glasgow, G12 8QQ, UK}




\runningtitle{Generalized regularization for solar bremsstrahlung X-ray spectra}
\runningauthor{Kontar et al.}

\begin{abstract}
Hard X-ray spectra in solar flares provide knowledge of the electron spectrum that results from
acceleration and propagation in the solar atmosphere. However, the inference of the electron
spectra from solar X-ray spectra is an ill-posed inverse problem. Here we develop and apply an
enhanced regularization algorithm for this process making use of physical constraints on the form
of the electron spectrum. The algorithm incorporates various features not heretofore employed in
the solar flare context : Generalized Singular Value Decomposition (GSVD) to deal with different
orders of constraints; rectangular form of the cross-section matrix to extend the solution energy
range; regularization with various forms of the smoothing operator; and ``preconditioning'' of the
problem. We show by simulations that this technique yields electron spectra with considerably more
information and higher quality than previous algorithms.
\end{abstract}
\end{opening}

\section{Introduction}
In order to address fundamental questions on electron propagation
and acceleration in solar flares, it is necessary to infer as much
as quantitative information as possible on the electron spectrum
in the solar plasma. A longstanding method of doing this involves
the analysis of the emitted hard X-ray (HXR) bremsstrahlung
spectrum, in particular the inversion of the integral equation
(Brown, 1971) relating the two spectra. This task is particularly
challenging since even the most accurate photon spectra are
contaminated by noise, which is dramatically amplified in any
unconstrained attempt to extract the electron flux spectrum.
Traditional approaches to the determination of the electron
spectrum sidestep this problem by assuming simple (e.g. isothermal
+ power-law) forms and adjusting their parameters to achieve the
best fit to the HXR data (e.g., Holman et al. 2003). However, such
algorithms, by their very nature, cannot detect features in the
electron spectrum that, although real, were not included into the
prescribed empirical form. Indeed, the analysis of high resolution
{\it RHESSI} (Ramaty High Energy Solar Spectroscopic Imager) (Lin
et al., 2002) spectra shows substantial deviations from simple
models (Kontar et. al., 2003). While this was adequate for earlier
low resolution data, the goal in high resolution photon spectrum
analysis should be the suppression of noise-induced unphysical
behaviour in the electron flux, while maintaining maximum ability
to recover faithfully real features. Various algorithms have been
employed with this intent (see, e.g., Johns and Lin 1992; Thompson
et al. 1992;  Piana 1994; Piana \& Brown 1998: Piana et al. 2003)
all belonging to the wide class of regularization methods for
linear ill-posed inverse problems, though differing in method of
regularization.  For example, Johns and Lin (1992) sum over energy
intervals to obtain sufficiently good statistical accuracy
(regularization by coarse energy binning). Their results suggested
downturn of the electron spectra below 50~keV, though the results
were too uncertain to be conclusive. Piana et al. (2003) have
detected, through Tiknonov regularized inversion (Tikhonov 1963),
a feature at $E \simeq 55$ keV in the mean source electron
spectrum for the July 23, 2002 solar flare, that has been
impossible to detect through a forward-fitting algorithm involving
power-law functions such as those used by Holman et al. (2003).
Although it has not been possible so far to establish the origin
of this particular feature \footnote{It is possible that this
particular feature has non-solar origin and is a result of the
effects of pulse pileup -- Smith et al. 2002}, nevertheless the
${\overline{F}}(E)$ form obtained by Piana et al. (2003) is a
faithful description of the ${\overline{F}}(E)$ corresponding to
the photon spectrum used.

An observed hard X-ray spectrum $I(\epsilon)$ is related, through
a bremsstrahlung cross-section $Q(\epsilon,E)$, to the {\it mean
electron flux spectrum} ${\overline{F}}(E)$ in the source, through
the relation (Brown 1971; Johns and Lin 1992; Brown, Emslie \&
Kontar 2003)
\begin{equation}
I(\epsilon) = {1 \over 4\pi R^2} \, \, \overline n V \int_\epsilon^\infty  \overline F(E) \,
Q(\epsilon,E)\, dE, \label{def}
\end{equation}
where $R$ is the distance to the observer, $V$ is the emitting
volume and ${\overline{n}}=V^{-1} \int n({\bf{r}}) dV$ is the mean
target density. The problem of determining ${\overline{F}}(E)$
from $I(\epsilon)$ when they are related by a Volterra-type
equation such as~(\ref{def}) is an ill-posed problem in the sense
of Hadamard (1923). As a result, every experimental problem
described by an equation like~(\ref{def}) is affected by a
numerical pathology termed {\it ill-conditioning} whereby, when an
unconstrained solution procedure is followed, the presence of
measurement noise is reflected in unphysical oscillations in the
reconstructed solution for the source function. To obtain a
meaningful solution ${\overline F}(E)$ one needs to avoid noise
amplification (e.g. Craig \& Brown 1986) by means of
regularization methods applying physical constraints to the
electron spectrum. The algorithm looks for an approximate
least-squares solution of the integral equation relating the
photon and the electron spectra, but subject to inclusion of
additional information based on physically meaningful constraints
or assumptions on the solution. This leads to the formulation of a
family of regularization methods exploiting different possible
{\em{a priori}} information on the electron flux coming from solar
physics. The aim of the present paper is to introduce this
generalized constraint approach into the field of solar HXR
spectrum analysis, together with various other new features
(physical, mathematical and numerical) yielding a more effective
algorithm,  based on Generalized Singular Value Decomposition, for
determination of ${\overline F}(E)$. More precisely, in this paper
we will discuss the three major problems concerning the
application of a regularization approach to the analysis of HXR
data, that is: how to effectively introduce physically meaningful
constraints into the inversion procedure; how to tune the
stability requirement avoiding artificial clustering in the
residuals, and, finally, how to adapt the regularization
techniques to solar data characterized by huge dynamical ranges.

\par

The plan of the paper is as follows. In \S2 we review the
mathematical formulation of the problem pointing out its numerical
instability. In \S3 we discuss the relation between regularization
and physical constraints, while in \S4 we describe the
regularization and the computational method based on the
Generalized Singular Value Decomposition. In \S5 a solution is
proposed, through analysis of the cumulative residuals,  to the
crucial problem of optimal choice of the regularization parameter
which controls the trade off between stability and loss of
information. In \S6 we perform an analysis of the solution
structure. Finally, in \S7 we present some applications in the
case of simulated photon spectra to demonstrate the improvements
achievable and the summary is presented in \S8.

In a subsequent paper (Kontar et al, 2004 (hearafter Paper II)) we
will apply this approach to real spectra observed with RHESSI (Lin
et al, 2002).

\section{Mathematical formulation}

Equation (\ref{def}) is a Volterra integral equation of the first kind and can be expressed in
terms of a linear integral operator $A:\mathbb{X} \rightarrow \mathbb{Y}$,
\begin{equation}\label{b1}
(A{\overline{F}})(\epsilon)\equiv\frac{1}{4\pi R^2} \, {\overline{n}} \, V
\int_{\epsilon}^{\infty} {\overline{F}}(E) \, Q(\epsilon,E) \, dE,
\end{equation}
where $\mathbb{X}$ and $\mathbb{Y}$ are two appropriate functional (Hilbert) spaces. For physical
forms of the bremsstrahlung cross-section $Q$, $A$ is a compact linear operator so that (Bertero
et al, 1985) every discretization of equation~(\ref{b1}) is characterized by (significant)
numerical instabilities. Let us, then, consider some convenient discretized form of~(\ref{def}),
viz. the linear system
\begin{equation}\label{b2}
{\bf{A}}{\bf{{\overline{F}}}} = {\bf{g}}~~~,
\end{equation}
where
\begin{eqnarray}\label{b3}
A_{ij} = \frac{{\overline{n}}V}{4\pi R^2}
Q((\epsilon_{i+1}+\epsilon _{i})/2,(E_{j+1}+E_{j})/2) \, \delta
E_j ~~~,~~~i=1,\ldots,N~~~j=1,\ldots,M (>N),
\end{eqnarray}
${\bf{{\overline{F}}}}=({\overline{F}}(E_1),\ldots,{\overline{F}}(E_M))$,
${\bf{g}}=(g(\epsilon_1),\ldots,g(\epsilon_N))$, with $M > N$, and
the $\delta E_j$ and $\delta \epsilon_i$ are appropriate weights.
The values $g(\epsilon_i)$ correspond to a set of discrete photon
counts in energy bands $\epsilon_i \rightarrow \epsilon_i + \delta
\epsilon_i$, while the ${\overline F}(E_j)$ are the corresponding
values of the mean electron flux in energy bands $E_j \rightarrow
E_j + \delta E_j$. We use a matrix ${\bf {A}}$ corresponding to
the bremsstrahlung cross-section due to Haug (1997) with the
Elwert (1939) Coulomb correction applied\footnote{Note that the
Elwert correction is not applicable for the high-frequency limit,
$\epsilon \simeq E$. However, since photons of energy $\epsilon$
are produced, in general, by a wide range of electron energies
$E$, the neglect of a more sophisticated form of the cross-section
in this very narrow energy range does not significantly affect our
results.}.

It is important to recognize that, since electrons of all energies
$E \geq \epsilon$ contribute to the photon emission at energy
$\epsilon$, in general the hard X-ray spectrum over a finite range
$[\epsilon_1,\epsilon_N]$ of photon energies contains information
on the {\it electron} spectrum over a much wider range, in
particular within the range $\epsilon_N < E < E_M$ above the
uppermost photon energy. For example, if ${\overline F}(E)$ has an
upper energy cutoff at $E=E_{\rm max} > \epsilon_N$, then this
cutoff will affect the observed photon spectrum below
$\epsilon_N$, since the photon spectrum must tend to zero at
$\epsilon \rightarrow E_{\rm max}$. Hence, by extending our array
of $E$ values to a sufficiently high value, the solution
of~(\ref{b2}) can potentially reveal evidence of an upper energy
cutoff (see \S7 below).

 Since $M>N$, problem (\ref{b2}) is underdetermined, i.e., there is no unique solution of the linear
system. The same holds true if we consider the least-squares problem
\begin{equation}\label{b4}
\|{\bf{A}}{\bf{\overline{F}}}-{\bf{g}}\|^2 = \min
\end{equation}
where $\|\cdot\|$ is the canonical Euclidean norm defined by
\begin{equation}
\|f\| \equiv \left( \int_{x_{min}}^{{x_{max}}}f^2(x)
\mbox{d}x\right)^{1/2} \label{norm}
\end{equation}
The solutions of (\ref{b4}) are commonly known as pseudosolutions
(Bertero et al, 1985). Obviously, additional constraints need to
be applied to obtain a unique solution.

 \section{Physical constraints and Regularization}

\hspace{6mm}
 As a physical quantity $\bf{\overline{F}}$ must satisfy various physical conditions
such as $\bf{\overline{F}}\ge 0$ and any known constraints such as properties which are to be
conserved or to be minimised/maximised. Many such properties can be expressed, for a suitable
closed operator $\bf L$, in the form
\begin{equation}\label{const}
\|{\bf{L\overline{F}}}\| \leq \mbox{const}
\end{equation}
leading to the need to solve least square problem (\ref{b4}) subject to additional constraint
(\ref{const}). This constrained minimum problem can be solved using the Lagrange multiplier
method, namely
\begin{equation}\label{mproblem}
\mathcal{L}(\bf{\overline{F}})\equiv \|{\bf{A}}{\bf{\overline{F}}}-{\bf{g}}\|^2+ \lambda
\|{\bf{L\overline{F}}}\|^2=\mbox{min}
\end{equation}
where $\lambda$ is a Lagrange multiplier. This approach to regularize the problem is known as
Tikhonov regularization (Tikhonov, 1963).

Three possible different choices for ${\bf L}$ are discussed in the following.

\par As a first example, we observe that the density-weighted target-averaged energy-integrated
flux of electrons is given by the Euclidean norm
$\|{\bf{\overline{F}}}\|$. Physically, this quantity must be fixed
by the total flux (the function of the total number of electrons
or energy), a requirement that can be met by a suitable choice of
the second (constraint) term in equation (\ref{mproblem}), the
simplest choice being $\|{\bf{\overline{F}}}\|$ or ${\bf L=1}$,
the identity matrix so that the second term of equation
(\ref{mproblem}) is just the Euclidean norm of the solution
${\bf{\overline{F}}}$. Problem (\ref{mproblem}) with this
constraint operator is also termed {\em{zero order
regularization}} and was used by Piana et al. (2003) for solar
data. It defines the non-parametric $\overline{F}(E)$ of
target-averaged total electron flux consistent with the data for a
given parameter $\lambda$ to be chosen via some appropriate
additional condition.

The source averaged electron flux ${\overline{F}}(E)$ physically results from a combination of the
injected electron flux $F_0(E_0)$ spectrum and the physics of particle transport in the radiating
source. Under a broad, but not comprehensive range of conditions, these flux spectra are related
by (Brown \& McKinnon, 1985)
\begin{equation}\label{ff0}
{\bf{\overline{F}}}(E) \sim \frac{1}{|dE/dN|} \int_{E}^{\infty} F_0(E_0)dE_0~~~,
\end{equation}
where $dE/dN$ is the rate of energy loss per unit column density . For example, for collisional
energy losses in a cold target, $dE/dN \sim -1/E$ and therefore (Brown and Emslie, 1988)
\begin{equation}\label{f0}
F_0(E_0) \sim - \frac{d}{dE} \left[\frac{{\bf{\overline{F}}}(E)}{E}\right]_{E=E_0}~~~.
\end{equation}
Equation (\ref{f0}) shows that, if the purpose of finding a
solution ${\overline{F}}(E)$ is to subsequently use that solution
to infer the injected electron flux spectrum $F_0(E_0) $, then the
mean electron flux should be differentiable, a requirement which
can be incorporated in (\ref{mproblem}) by adopting for ${\bf L }$
the differentiation operator ${\bf L} \sim {\bf D }^1$ which is
termed {\it first order regularization}.

Physically, if the electron acceleration can be described
deterministically, e.g. the injection function resulted from
acceleration can be presented similar to an integral (\ref{ff0})
with systematic acceleration instead of deceleration, then the
injected spectrum should be a differential function. For example,
acceleration by an electric field leads to differentiable spectrum
of accelerated (injected) electrons. If we believe the injection
function is differentiable, then ${\overline{F}}(E)$ should not
only be differentiable, but have a differentiable first
derivative. This corresponds to the requirement of a bounded
second order derivative, hence to {\it second order
regularization} ${\bf{L}} \sim {\bf{D}}^2$.

It is informative to consider the effect of  applying regularization methods of very high order,
since they can obscure physical features in ${\overline{F}}(E)$ that do not comply with the
imposed smoothness constraint. The $k$-order derivative ${\bf{D}}^{(k)} {\bf{\overline{F}}}$
cannot be defined over less than $k$ points, which reduces the dimension of the solution space to
$N-k$ for $N$ data points. This makes higher order solutions more restrictive and potentially less
precise though in practice the high resolution data now available have such large $N\sim 300$ that
this is not an issue for any reasonable order of regularization. However in the case of
forward-fitting procedures, the natural $N\sim 300$ dimensional space of  possible solutions is
forcefully squeezed into the space of only  a few dimensions ( 5 in the case of power-law +
isothermal components - Holman et al, 2003) which corresponds to very high order regularization
and hence is very restrictive.

\section{Regularized solution and Generalized Singular Value Decomposition}

Provided that the null spaces of the matrices ${\bf{A}}$ and ${\bf{L}}$ intersect trivially (i.e.
${\bf{AF}}={\bf 0}$ and ${\bf LF}={\bf 0}$ have no common solutions other than ${\bf F=0}$), the
formal solution of the minimum problem can be shown to be (Hansen, 1992)
\begin{equation}\label{solution}
{\bf{\overline{F}}}_{\lambda} = ({\bf{A}}^{*}{\bf{A}}+\lambda {\bf{L}}^{*} {\bf{L}})^{-1}
{\bf{A}}^{*}~~~.
\end{equation}
where ${\bf{A}}^{*}$ is adjoint of an operator ${\bf{A}}$.
 From the point of view of applications, this formula is not helpful
since truncation errors imply a notable loss of information in
forming the cross-product matrix ${\bf{A}}^{*}{\bf{A}}$ and,
furthermore, the computational effort required is significant.
Computational heaviness, together with the presence of local
minima, affect also the use of quadratic programming for convex
functional minimization, which is a typical strategy for computing
the solution of (\ref{mproblem}). A more effective approach is to
use Generalized Singular Value Decomposition (GSVD) algorithm.
Following van Loan (1976), we consider an $M\times N$ matrix ${\bf
A}$ and a $P\times N$ matrix ${\bf L}$ ($M\geq N\geq P$). Then for
each pair of real matrices ({\bf A},{\bf L})
\begin{equation}
{\bf A}\in \mathbb{R}^{M\times N}, \quad {\bf L}\in \mathbb{R}^{P\times N}. \label{rank}
\end{equation}
it can be shown that there exists a set of singular values $\sigma^A_k$, $\sigma^L_k$ satisfying
the relation $(\sigma^A_k)^2+(\sigma^L_k)^2=1$, and singular vectors $\tilde{{\bf u}}_k,
\tilde{{\bf v}}_k, \tilde {\bf w}_k$, where the first two sets are orthogonal and the third one is
a set of linearly independent vectors satisfying the simultaneous equations

\begin{eqnarray}
{\bf A}=\tilde{{\bf U}}\left(\begin{array}{lr}
               \mbox{diag($\sigma ^A_k$)},&\mbox{$0$}\\
               \mbox{$0$} &\mbox{${\bf 1}_{N-P}$}\\
               \mbox{$0$} &\mbox{$0$}
               \end{array}
\right) {\tilde {\bf W}}^{-1}, \;\;\; {\bf L}=\tilde{{\bf V}}(\mbox{diag($\sigma ^L)$}\;\; 0)
{\tilde {\bf W}}^{-1}. \label{gsvd}
\end{eqnarray}
  Here the $M \times M$ matrix $\tilde{{\bf U}}$ is formed from the
$M$ column vectors $\tilde{{\bf u}}_k, k = 1,\ldots, M$, with similar definitions for the $P
\times P$ matrix $\tilde{{\bf V}}$ and the $N \times N$ matrix $\tilde{{\bf W}}$. The generalized
singular values are defined as the ratios $\sigma _k=\sigma^A_k/\sigma^L_k$.

The solution to this generalized minimization problem~(\ref{mproblem}) can be shown to be (Hansen
1992)
\begin{equation} {\overline {\bf
F}}_{\lambda} = \sum_{k=1}^P \left({\sigma_k^2 \over \sigma_k^2 + \lambda} {( {\bf g} \cdot
{\tilde {\bf u}}_k )\over \sigma^A_k}\right) \, {\tilde {\bf w}}_k+\sum_{k=P+1}^N ( {\bf g} \cdot
{\tilde {\bf u}}_k ){\tilde {\bf w}}_k. \label{soln}
\end{equation}
In the particular case when ${\bf L}={\bf 1}$ (zero-order regularization), $P=N$ and
Equation~(\ref{gsvd}) shows that ${\tilde {\bf u}}_k = {\bf u}_k, {\tilde {\bf v}}_k = {\bf v}_k$.
Hence the second term in the solution~(\ref{soln}) vanishes identically and the first term reduces
to the one for the zero-order regularized solution.

\section{Choice of the Regularization Parameter}

It has to be recalled that the choice of the regularization
parameter $\lambda$, and indeed of $\bf L$, has to be made
independently of the equation and of the data, using prior
knowledge or prejudice, since there is no unique solution to the
equation (\ref{b4}) itself. As mentioned, the integral properties
of the electron flux (\ref{const}), if they were known, would
unambiguously determine the Lagrange multiplier $\lambda$ in our
problem (\ref{mproblem}). Unfortunately, we do not know {\it a
priori} the total flux of X-ray producing electrons or other
integral quantity. Therefore it is advantageous to use knowledge
of the errors in the recovered solution to choose the
regularization parameter. Several criteria for determining the
optimal $\lambda$ in Equation~(\ref{soln}) have been introduced. A
general property of them is that the optimal $\lambda$ tends to
zero when the noise level tends to zero. For example, according to
the {\it discrepancy principle} (Tikhonov et al, 1995), the best
value of $\lambda$ is given by the solution of the equation
\begin{equation}\label{discrepancy}
\|{\bf{A}}{\overline{\bf{F}}}_{\lambda} - {\bf{g}}\|^2 = \|\delta {\bf g}\|^2,
\end{equation}
where $\delta g$ is some measure of the noise affecting the data (essentially, typically the
canonical norm of the error vector). The discrepancy principle is typically rather robust in that
it yields stable values but empirical tests show that the parameter it provides is often too
large, and the corresponding regularized solution oversmoothed.

Here we propose a procedure for the choice of the regularization parameter based on the analysis
of the residuals $r_k=(({\bf{A}}{\overline{\bf{F}}})_k - {\bf{g}}_k)/\delta g_k$. Then the
deviation weighted by the error
\begin{equation}\label{chi2}
\|({\bf{A}}{\overline{\bf{F}}}_{\lambda} - {\bf{g}})(\delta {\bf g})^{-1}\|^2 \simeq 1
\end{equation}
accounts more accurately for point-to-point error variation than
(\ref{discrepancy}). Indeed, $\lambda $ defined by Equation
(\ref{chi2}) has accounted for detailed structure of errors, while
the discrepancy principle uses only total error.

Ideally, the normalized residuals should be consistent with statistical deviations in the data,
and should therefore form a gaussian distribution with zero mean; the cumulative normalized
residual
\begin{equation}
C_j = \frac{1}{j} \, \sum_{k=1}^j {r_k} \label{cum}
\end{equation}
should be mostly within $\pm 1/\sqrt{j}$. A too-small set of values for the cumulative residuals
indicates insufficient smoothing, whereas a set of values that consistently exceed $\pm
1/\sqrt{j}$ (especially if the sign of the residuals cluster) indicates too great a regularizing
parameter. Therefore, we start with the value of $\lambda$ given by (\ref{chi2}) and then reduce
$\lambda$ until the average residual in the photon spectrum over the energy range from
$\epsilon_1$ to $\epsilon_j$ as a function of $j$ is mostly within the $\pm \sigma$ limits
expected if the residuals were purely statistical, drawn from a normal distribution. This
technique is similar to requiring $\chi^2 \simeq 1$ in hypothesis testing situations.

\section{Analysis of the solution structure}

\begin{figure}
\includegraphics[width=89mm]{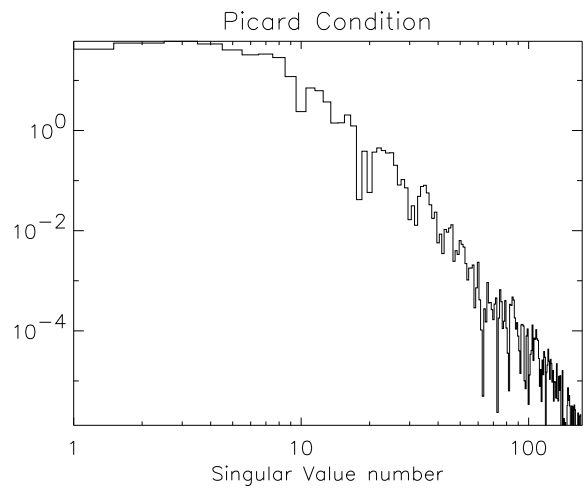}
\caption{Variation of the coefficients $c_k$ in Equation (\ref{ck}) as a function of the vector
number $k$ for the simulated data set with $\delta=2$.} \label{picard}
\end{figure}

A key issue in achieving the most meaningful construction of a regularized solution lies in
analysis of the quantities
\begin{equation}\label{ck}
c_k = \left \vert {(\sigma_k)^2 \over (\sigma_k)^2 + \lambda} {( {\bf g} \cdot {\tilde {\bf u}}_k
)\over \sigma^A_k} \right \vert~~~~k=1,\ldots,N
\end{equation}
i.e., the absolute value of the coefficients which multiply the singular vectors ${\bf{v}}_k$ in
the solution~(\ref{soln}). To obtain a meaningful solution represented by Equation~(\ref{soln})
the singular values $\sigma_{k}$ should decrease faster on average than the coefficients
$({\bf{g}},{\bf{u}}_k)$ (the {\it Picard condition}; Groetsch 1984). Figure ~\ref{picard} shows a
typical behaviour of the coefficients $c_k$.

The first singular vector is always either positive or negative definite (so that $c_1 {\tilde
{{\bf w}_1}}$ is always positive), while the ${\tilde {\bf{w}}}_k$ always have oscillatory
behavior for $k>1$. Therefore, for sufficiently small regularization parameters $\lambda$, the
regularized solution may exhibit negative values when the coefficients $c_k$ are not decreasing
fast enough, so that relatively large values of the regularization parameter $\lambda$ are
necessary to guarantee a positive definite solution.

Inasmuch as typical solar flare hard X-ray spectra are
sufficiently steep, it is advantageous to transform the
fundamental problem~(\ref{def}) (or,
equivalently,~[\ref{mproblem}]) to a form in which the $c_k$ are
decreasing faster with $k$, so that smaller values of $\lambda$,
which preserve more fidelity in the recovered solution, can be
used. This will also avoid errors connected with finite precision
arithmetics in machine calculations.

Two strategies can be followed to overcome this difficulty. In the first one, instead of
considering equation (\ref{b2}) we consider the new linear system
\begin{equation}\label{aref}
{\bf{A}} \delta {\overline{\bf{F}}} = \delta {\bf{g}}
\end{equation}
with
\begin{equation}\label{fref}
\delta {\overline{\bf F}} = {\overline{\bf F}} - {\overline{\bf F}}_*; \quad \delta {\bf g} = {\bf
g} - {\bf A} {\overline{\bf F}}_*,
\end{equation}
where ${\overline{\bf F}}_*$ is an adopted  form (often, but not
necessarily, a closed parametric expression) for
${\overline{\bf{F}}}$.  Now $\delta {\bf{g}}$ represents the
deviations from the reference spectrum.  Hence, if this reference
spectrum is chosen appropriately (say from a forward-fit solution
-- e.g., Holman et al. 2003), then these deviations $\delta
{\overline{\bf F}}$ vary around zero, and the function $\delta
{\bf{g}}$ will be significantly flatter than ${\bf g}$, so that
the behavior of the $c_k$ becomes more monotonic.

In addition to facilitating the calculation of a smooth (but not unnecessarily oversmoothed)
solution of the minimization problem, the quantity $\delta {\overline{\bf F}}$ is interesting in
its own right. It represents the deviation of the actual electron spectrum ${\overline{\bf F}}
(E)$ from the assumed reference spectrum ${\overline{\bf F}}_* (E)$ and hence is an ``adjustment''
to this assumed (e.g., forward-fitted) form. This "initial guess" approach is the one adopted in
the present paper.

A second possible strategy to constrain the behaviour of the $c_k$ is to consider the change of
variables
\begin{equation}\label{scal}
g(\epsilon_i) \rightarrow \epsilon_i^p \, g(\epsilon_i); \quad {\overline F}(E_j) \rightarrow
E_j^q \, {\overline F}(E_j); \quad {\bf A}_{ij} \rightarrow {\epsilon_i^p \over E_j^q} \, {\bf
A}_{ij},
\end{equation}
with $p,q$ positive real numbers. Then the basic form of the solution~(\ref{soln}) is unaltered,
but the forms of the matrix and its associated singular system are altered which leads to modified
behavior of the coefficients $c_k$. If a scaling~(\ref{scal}) can be found that makes the $c_k$
decreasing functions of $k$, then this transformed solution will have more desirable properties.
We have found through experimentation that a judicious choice of scaling is $p=q=(\gamma-1)/2$,
where $\gamma$ is the best-fit power-law spectral index to the array of $g$ values. For all values
of $\gamma$ (from 2 to $\approx 20$), this scaling drives the coefficients $c_k$ toward a rapidly
decreasing form. Note that while re-scaling with $p=\gamma$ produces a much flatter data function
${\bf {g}}$, such a steeper scaling concomitantly leads (equation~[\ref{scal}]) to the matrix
${\bf {A}}$ becoming less diagonal, thereby increasing the ill-conditioning of the system and
hence the rate of decay of the $\sigma_k$ with $k$. The choice $p = (\gamma-1)/2$ hence lies at
the ``middle ground'' between steepness of the input data vector and ill-conditioning of the
transformation matrix; similar arguments apply to the choice of $q$.

It should also be noted that ``rescaling'' is equivalent to constructing the {\it ratio} (rather
than the {\it difference}) of ${\overline F}(E)$ relative to a reference ${\overline F}_*(E)$
form.

\section{Application of the Algorithm to Simulated Data}

In this section we explore the application of the above techniques to simulated data, in order to
demonstrate some of the effects of various features.

\subsection{Non-square Matrix}
\begin{figure}
\begin{center}
\includegraphics[width=89mm]{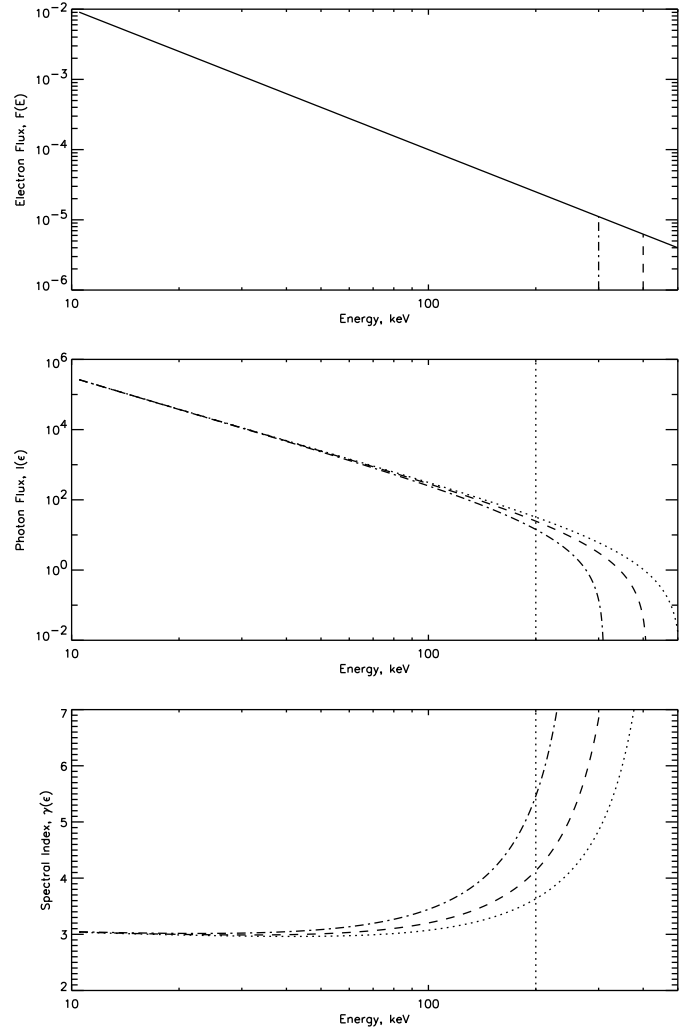}
\end{center}
\caption{{\it Upper panel:} Simulated mean source electron spectra ${\overline F}(E)$,
which has the form of a power-law (index $\delta=2$), with three different upper energy cutoffs
$E_{\rm max}$.  {\it Middle panel:} Corresponding photon spectra $I(\epsilon)$ for each of the
${\overline F}(E)$ forms.  {\it Lower panel:} Local spectral index $\gamma = d \log I(\epsilon)/ d
\log \epsilon$ for each photon spectrum.  Note the evidence for the high-energy cutoff in the hard
X-ray spectrum and its local spectral index at much lower energies than $E_{\rm max}$.  The
vertical lines at $\epsilon = 200$~keV represent the upper energy limit of the spectral data used
for subsequent analysis.}\label{elec_phot_spec}
\end{figure}

To demonstrate the benefits of using a non-square $\bf A$ extending to higher electron energies
than the highest observed photon energy, we assumed an {\it actual} ${\overline F}(E)$ of the form
${\overline F}(E) \sim E^{-\delta}, 10 < E < E_{\rm max}$, with $\delta =2 $ and $E_{\rm max} =
(300, 400, 500)$~keV.  We then used Equation~(\ref{def}) and the electron-proton bremsstrahlung
cross-section from Haug (1997) to calculate the photon spectrum $I(\epsilon)$, in the range $10
\leq \epsilon \leq 200$~keV, from each electron spectrum. (It should be noted that the upper limit
to the ``observed'' photon spectrum is in all cases less than the upper-energy cutoff in the
electron spectrum.) Figure~\ref{elec_phot_spec} shows the calculated photon spectra and the
corresponding local spectral index $\gamma = - d \log I(\epsilon)/ d \log \epsilon$. At low
energies, the spectrum is well represented by a power-law form with $\gamma = \delta + 1 =3$, but
even at energies as low as $\sim 50$~keV, the deviation from a power-law behavior induced by the
upper-energy cutoff in ${\overline F}(E)$ is already evident. For the case $E_{\rm max} = 300$~keV
, by $\epsilon = 100$~keV the spectrum has steepened from its low-energy form ($\sim
\epsilon^{-3}$) sufficiently that the spectral index has increased by as much as 0.5.

Such a deviation in local hard X-ray spectral index $\gamma$ is clear evidence for a significant
deviation from the power-law behavior of the generating ${\overline F}(E)$ spectrum at higher
energies, although the exact nature of this high-energy deviation is not immediately obvious from
the form of the photon spectrum (or even from a $\gamma [\epsilon]$ plot). We therefore now
explore the ability of our technique to uncover the actual nature of this deviation (namely, in
this case the upper energy cutoff at $E=E_{\rm max}$). We performed a regularized inversion of the
photon spectra of Figure~\ref{elec_phot_spec} by using solution (\ref{soln}) under the following
conditions: zero order regularization; the simulated photon ``data'' used covered the range $10 <
\epsilon < 200$~keV and the electron upper energy limits $E_{\rm upper}$ used in the inversion
were 400, 500, and 600~keV.

\begin{figure}
\begin{center}
\includegraphics[width=89mm]{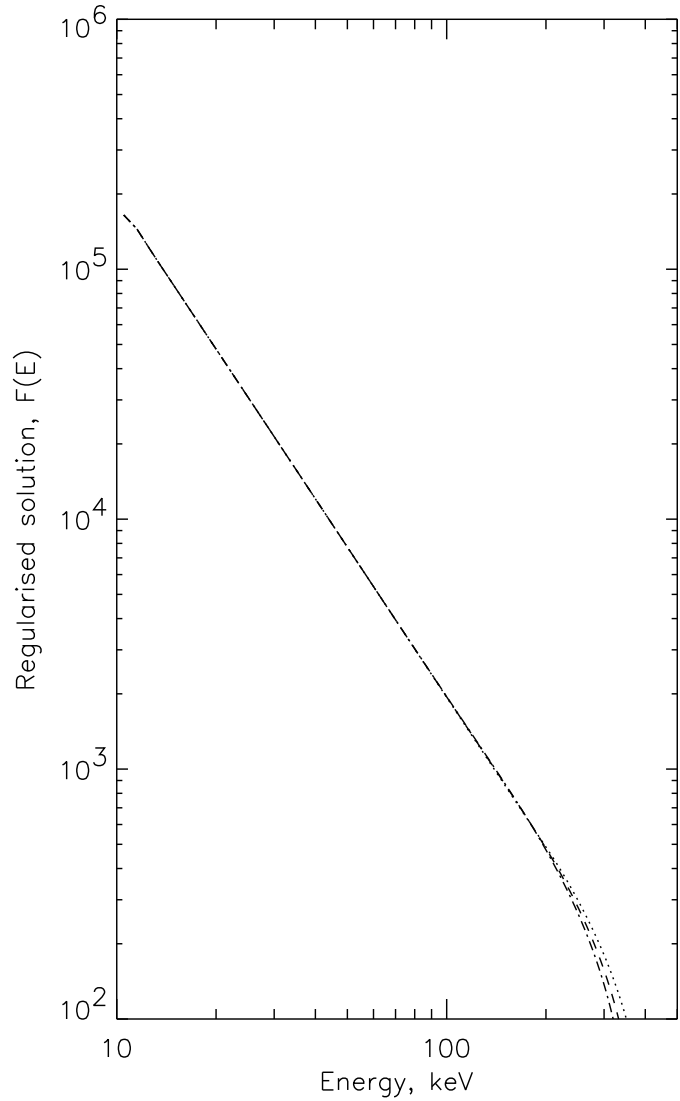}
\end{center}
\caption{Mean source electron spectra ${\overline F}(E)$ recovered from the photon
spectrum $I(\epsilon; E_{\rm max} = 300$~keV) (dot-dashed spectrum in
Figure~\ref{elec_phot_spec}), using the GSVD technique and zero order regularization with photon
``data'' in the range $10 < \epsilon < 150$~keV and electron energy ranges $10$~keV~$< E < E_{\rm
upper}$, where $E_{\rm upper} = 400$~keV (dot-dashed curve), 500~keV (dashed curve) and 600~keV
(dotted curve). The reference spectrum (Equation~[\ref{fref}]) was of the form ${\overline F}_*(E)
\sim E^{\gamma_{*}}$. Note that the high-energy behavior of ${\overline F}(E)$ (in particular the
high-energy cutoff at $E_{\rm max} = 300$~keV) is quite faithfully reproduced.} \label{sim_recov}
\end{figure}

\begin{figure}
\begin{center}
\includegraphics[width=89mm]{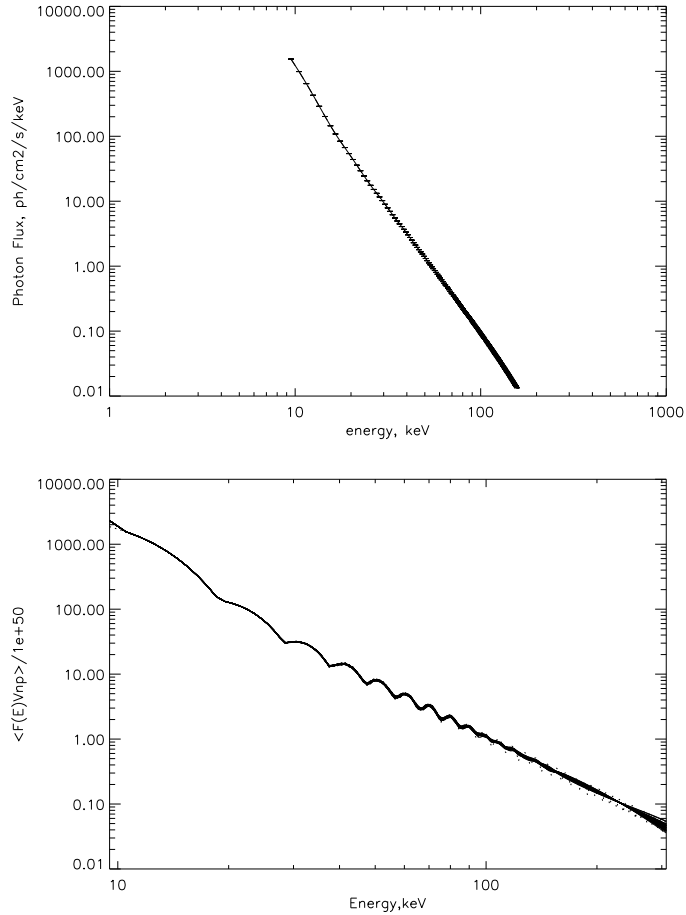}
\end{center}
 \caption{Reconstruction from simulated data: upper panel: the
simulated photon spectrum; bottom panel: the reconstruction obtained by using first order
regularization. The dash line shows the input electron spectrum and solid lines presents 30
realizations of the solution with the data randomly perturbed within error bars.}\label{sim_sin}
\end{figure}

In all cases the recovered electron spectra quite faithfully reproduce the actual high-energy
cutoffs $E_{\rm max} = 300$~keV (Figure~\ref{sim_recov}). The higher energy cut-off is seen as
substantial steepening in the reconstructed spectra. However, we should note somewhat obvious
limitations: if the true spectra has some variations above the range we are given, the
reconstructed solution does not display this features due to the lack of information given
(Figure~\ref{sim_sin}).

\subsection{Higher order regularization with GSVD}

To illustrate the benefits of GSVD and higher order regularization
we consider the reconstruction of an electron spectrum which is
the superposition of a power-law plus an oscillatory trigonometric
function. In this case, the use of a first order penalty term
governed by ${\bf{L}} \sim {\bf{D}}^1$ is more effective than the
zero-order regularization algorithm, since the bound on the first
derivative of the regularized solution assures a sufficiently
correct behaviour (in terms of residuals) of the solution. In
Figure 4 (upper panel) the simulated photon spectrum is obtained
by inserting the theoretical electron spectrum into equation
(\ref{b2}) while Figure 4 (lower panel) shows the reconstruction
obtained by using formula (\ref{soln}) when ${\bf{L}}\sim
{\bf{D}}^1$ and $\lambda$ is chosen by using the cumulative
residuals analysis criterion (in this figure we also superimposed
the theoretical electron spectrum in order to point out the
reliability of our approach). In Figure 5(upper panel) and Figure
5(lower panel) the behaviour of the normalized and cumulative
residuals shows that the fitting performance of the regularized
solution is extremely accurate in the range below $150$ keV.

The importance of higher order regularization using GSVD also
becomes explicit when one wants to derive the injected electron
distribution $F_0(E_0)$. Figure (\ref{Fbar}) shows both the mean
source and injected electron spectra obtained using different
orders of regularization, compared with their true forms. As a
true form of $F_0(E_0)$ we took a power-law with a bump simulated
by exponent as can be seen in Figure (\ref{Fbar}). The simulated
photon spectrum is obtained by inserting the theoretical electron
spectrum with the high energy cut-off at 300 keV into equation
(\ref{b2}) and $5$\% noise has been added. The reconstructed
spectra and input spectra are shown in the Figure (\ref{Fbar}).
Note, that in the reconstruction we used data only up to 150 keV.

The second order regularized solution shows the closest solution for $F_0(E_0)$ in equation
(\ref{Fbar}). On the other hand first and zero order regularizations show results with smaller
$\chi ^2$ in recovering the spatially integrated spectra (\ref{Fbar}). Indeed, second and first
order regularization preserves information on small scale (closer reproduction of a hump in
$F_0(E_0)$ around $30$~keV), while zero order regularization is better in global properties of the
solution. The first and zero order regularization show lack of sufficient smoothness that is
displayed in oscillations in $F_0(E_0)$ (\ref{Fbar}). The main deviations from the true solutions
are observed above 200 keV, where we have only approximate solution and near low energy cut-off
due to boundary effects.

\section{Conclusions}
In the paper, we have summarized the essential mathematics associated with application of an
Generalized Singular Value Decomposition (GSVD) technique to the solution of Volterra integral
equations arising in solar X-ray spectroscopy, and in particular to the inference of mean source
electron spectra ${\overline F}(E)$ and injected (accelerated) electron spectrum $F_0(E_0)$ from
observations of solar flare hard X-ray spectra $I(\epsilon)$. Judicious use of this methodology
can recover forms of ${\overline F}(E)$ that are not only relatively free from the effects of data
noise amplification, but which also recover features that are not realizable using more
traditional (e.g., forward-fitting) methods. Further, they can reveal approximate behaviour in the
electron spectrum well above the range of photon energies observed.

In the companion paper (Paper~II) we will make use of these techniques in the analysis of
high-resolution solar flare spectra observed by {\it RHESSI}.

\acknowledgements This work was supported by NASA's Office of Space Science through Grants
NAG5-207745, by a PPARC Rolling Grant and by a collaboration grant from the Royal Society.

\begin{figure}
\begin{center}
\includegraphics[width=89mm]{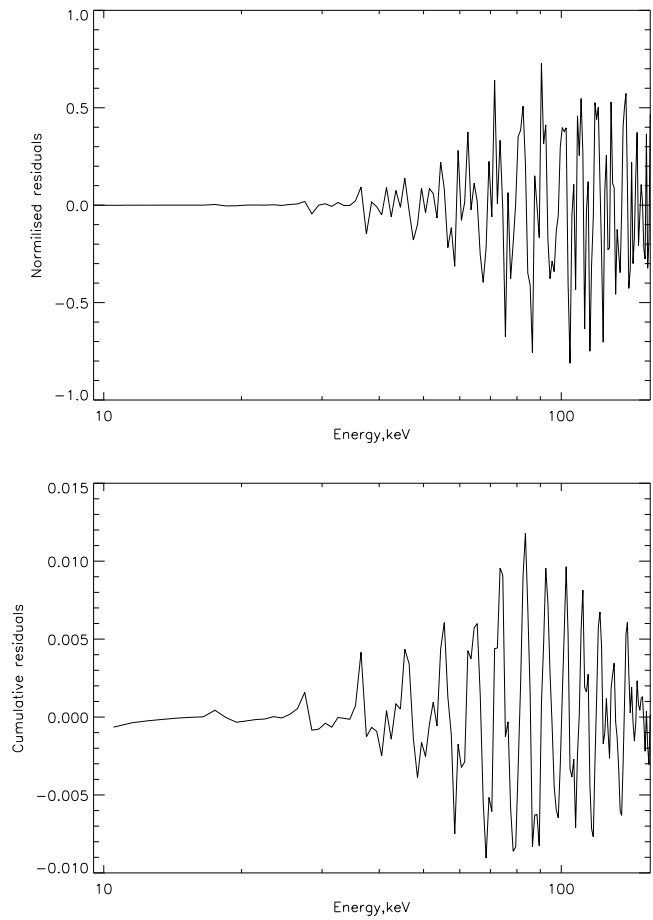}
\end{center}
\caption{Reconstruction from simulated data: upper panel: normalized residuals; bottom
panel: normalized cumulative residuals}
\end{figure}

\begin{figure}
\begin{center}
\includegraphics[width=89mm]{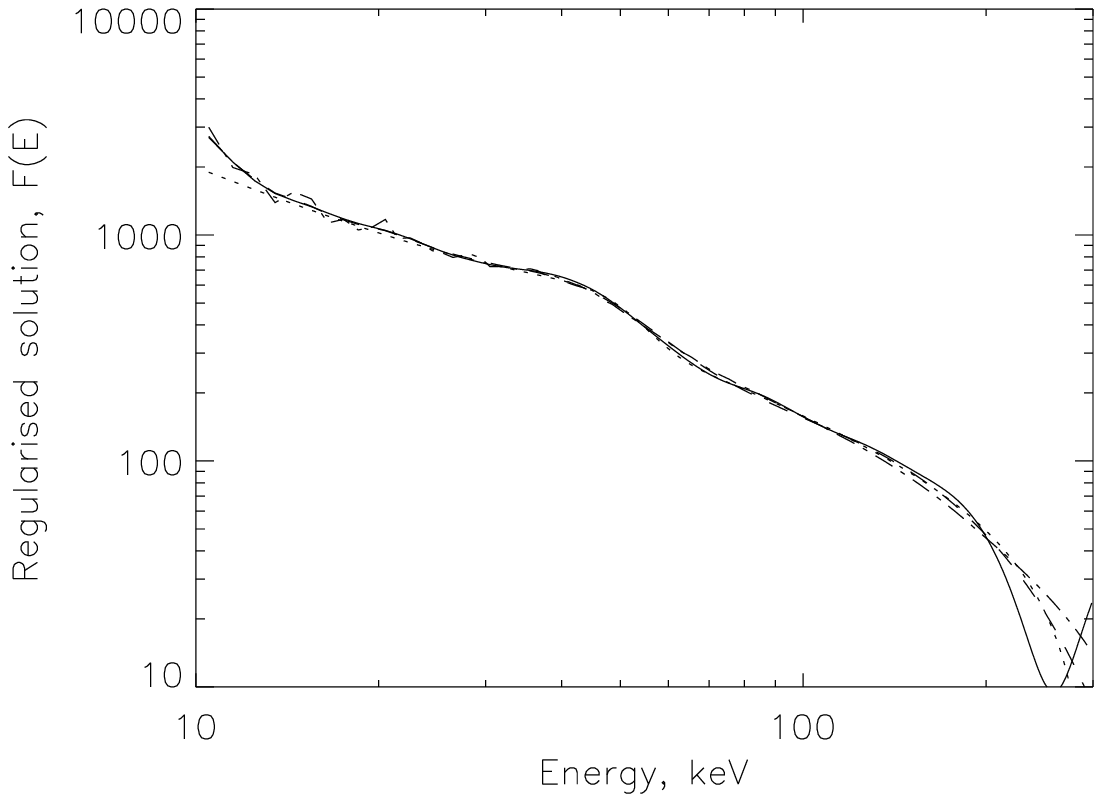}
\includegraphics[width=89mm]{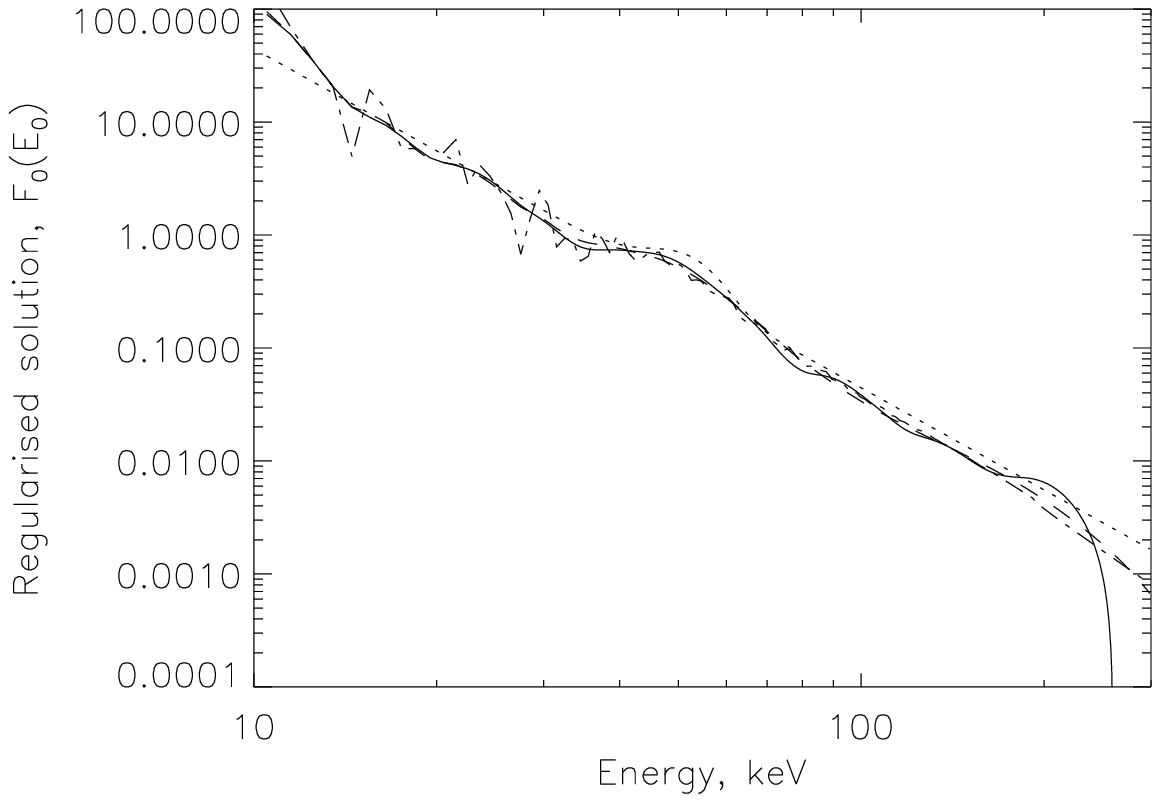}
\end{center}
 \caption{Reconstruction from simulated data. Recovered mean flux ${\overline F}(E)$(upper panel) and recovered injected flux $F_0(E_0)$ (lower panel)
 given by equation (\ref{f0}). The dotted lines shows the true solution, while three various orders of regularization
are shown: second order regularization (solid line), first order (dash line), zero (dash dot).}
\label{Fbar}
\end{figure}

\end{article}


\begin{thebibliography}{}

\bibitem[]{} Bertero,M., De Mol,C., and Pike,E.R.: 1985, {\it Inverse Problems,} {\bf 1,} 300

\bibitem[]{} Bertero,M., De Mol, C., and Pike,E.R.: 1988, {\it Inverse Problems,} {\bf 4}, 573

\bibitem[]{} Brown, J.C., \& MacKinnon, A.L.: 1985, {\it Astrophys. J.}, {\bf 292}, L31

\bibitem[]{} Brown, J.C., \& Emslie, A.G.: 1988, {\it Astrophys. J.}, {\bf 331}, 554

\bibitem[]{} Brown, J.C., Emslie, A.G., \& Kontar, E.P.: 2003, {\it Astrophys. J. Lett.}, {\bf 595}, L115

\bibitem[]{} Craig, I.J.D. \& Brown, J.C.: 1986, {\it Inverse Problems in Astrophysics},
Bristol:Adam-Hilger.

\bibitem[]{} Elwert, G.: 1939, Ann. Physik, {\bf 34}, 178

\bibitem[]{} Hadamard, J.: 1923, {\it Lecture's on Cauchy's problem in linear partial differential equations},
Yale University Press: New Haven.

\bibitem[]{} Emslie, A.G., Kontar, E.P., Krucker, S., \& Lin, R.P.: 2003, {\it Astrophys. J. Lett.}, {\bf 595}, L107

\bibitem[]{} Groetsch, C.W.: 1984, {\it The Theory of Tikhonov Regularization for Fredholm
Equations of the First Kind}, (Pitman: Boston)

\bibitem[]{} Hansen, P.C.: 1992, {\it Inverse Problems}, {\bf 8}, 849

\bibitem[]{} Haug, E.: 1997, {\it Astron. Astrophys.}, {\it 326}, 417

\bibitem[]{} Holman, G. D., Sui, L., Schwartz, R. A., \& Emslie, A. G.: 2003, {\it Astrophys. J. Lett.}, {\bf 595}, L97

\bibitem[]{} Johns, C., \& Lin, R.P.: 1992, {\it Solar Phys.}, {\it 137}, 121

\bibitem[]{} Kontar, E.P., Brown, J.C., Emslie, A.G., Schwartz, R.A., Smith, D.M., and Alexander,
R.C.: 2003, {\it Astrophys. J. Lett.}, {\bf 595}, L123.

\bibitem[]{} Kontar, E.P., Piana, M., Emslie, A.G., \& Brown, J.C.: 2004, {\it Solar
Physics}, this volume (Paper II).

\bibitem[]{} Lin, R.P., et al.: 2003, {\it Astrophys. J. Lett.}, {\bf 595}, L69

\bibitem[]{} Lin, R.P., et al.: 2002, {\it Solar Physics}, {\bf 210}, 3

\bibitem[]{} Piana, M.: 1994, {\it Astron. Astrophys.}, {\bf 288}, 949

\bibitem[]{} Piana, M. \& Brown, J.C.: 1998, {\it Astron. Astrophys.}, {\bf 132}, 291

\bibitem[]{} Piana, M., Massone, A.M., Kontar, E.P., Emslie, A.G., Brown, J.C., \& Schwartz, R.A.:
2003, {\it Astrophys. J. Lett.}, {\bf 595}, L127

\bibitem[]{} Smith, D.M., et al.: 2002, {\it Solar Physics}, {\bf 210}, 33

\bibitem[]{} Thompson A.M., Brown J.C., Craig, I.J.D., \& Fulber, C.: l992, {\it Astron. Astrophys.}, {\bf 265}, 278

\bibitem[]{} Tikhonov, A.N.: 1963, {\it Sov. Math. Dokl.}, {\bf 4}, 1035

\bibitem[]{} Tikhonov, A.N., Goncharsky, A., Stepanov, V., and Yagola, A.:
1995, {\it Numerical methods for the solution of ill-posed
problems}, Kluwer: Dordrecht

\bibitem[]{} van Loan, C.F.: 1976, {\it SIAM J. Num. Anal.}, {\bf 13}, 76

\end{thebibliography}
\end{document}